\documentclass[english]{article}
\usepackage[T1]{fontenc}
\usepackage[latin9]{inputenc}
\usepackage{subfigure,geometry}
\usepackage{xcolor}
\usepackage{amstext}
\usepackage{amssymb}
\usepackage{stmaryrd}
\usepackage{stackrel}
\usepackage{graphicx}
\usepackage{esint,cite}

\makeatletter

\providecommand{\tabularnewline}{\\}

\makeatother

\usepackage{babel}
\begin{document}
\title{Generalized binomial state: Nonclassical features observed through
various witnesses and a measure of nonclassicality}
\author{Kathakali Mandal{\normalsize{}$^{\mathsection}$}, Nasir Alam{\normalsize{}$^{\dagger}$},
Amit Verma{\normalsize{}$^{\mathsection}$}, Anirban Pathak{\normalsize{}$^{\dagger}$}and
J.  Banerji$^{*}$\\
{\normalsize{}$^{\mathsection}$}Jaypee Institute of Information Technology,
Sector-128, Noida, UP-201304, India\\
{\normalsize{}$^{\dagger}$}Jaypee Institute of Information Technology,
A-10, Sector-62, Noida, UP-201309, India\\
$^{*}$Physical Research Laboratory, Navrangpura, Ahmedabad-380 009,
India}
\maketitle
\begin{abstract}
Experimental realization of various quantum states of interest has become possible in the recent past due to the rapid developments in the field of quantum state engineering. Nonclassical properties of such states have led to various exciting applications, specifically in the area of quantum information processing. The present article aims to study lower- and higher-order nonclassical features of such an engineered quantum state (a generalized
binomial state based on Abel's formula). Present study has  revealed that
the state studied here is highly nonclassical. Specifically, higher-order nonclassical properties of this state are reported using a set
of witnesses, like higher-order
antibunching, higher-order sub-Poissonian photon statistics, higher-order
squeezing (both Hong Mandel type and Hillery type). A set of other
witnesses for lower- and higher-order nonclassicality (e.g., Vogel's
criterion and Agarwal's A parameter) have also been explored. Further, an analytic expression for the Wigner function of the generalized binomial state is reported and the same is used  to witness
nonclassicality and to quantify the amount of nonclassicality
present in the system by computing the nonclassical volume (volume
of the negative part of the Wigner function). Optical tomogram of the generalized binomial state is also computed for various conditions  as Wigner function cannot be measured directly in an experiment in general, but the same can be obtained from the optical tomogram with the help of Radon transform. 
\end{abstract}

\section{Introduction}

With the advent of quantum state engineering \cite{marchiolli2004engineering,miranowicz2004dissipation,vogel1993quantum,sperling2014quantum}
and quantum computation \cite{nielsen2002quantum,pathak2013elements},
much attention has been given to the nonclassical properties of quantum
states \cite{barnett2018statistics,verma2010generalized,pathak2014wigner,verma2008higher,fu1997hypergeometric,moussa1998generation}.
The reason behind this intense attention is obvious as the  nonclassical states being the states having no classical analogue must be essential for  performing  tasks that are impossible in the classical
world (e.g., teleportation, densecoding, unconditionally secure quantum key distribution). In other words, nonclassical states which are characterized
by the negative values of Glauber-Sudarshan $P$-function, can only
establish quantum supremacy \cite{harrow2017quantum,neill2018blueprint}. 

Well known nonclassical properties are squeezing, antibunching and
entanglement. These nonclassical features have been reported in various
physical systems including optical couplers \cite{thapliyal2014higher,thapliyal2014nonclassical},
Bose-Einstein condensates \cite{giri2014single}, optomechanical systems
\cite{alam2017lower,alam2015approximate,alam2016nonclassical}, and
in many families of quantum states \cite{alam2018higher,Meher2018,malpani2018lower,Alam2018}.
Further, applications of squeezed states are known in the context
of LIGO experiment (which has been used successfully to detect gravitational wave)
\cite{abbott2016bp,abbott2016gw151226}, continuous variable quantum
key distribution \cite{gottesman2003secure,cerf2001quantum,madsen2012continuous,weedbrook2012gaussian};
application of entanglement is known in the context of quantum teleportation
and quantum cryptography \cite{bennett1993teleporting,ekert1991quantum,bennett1992quantum},
and antibunching is known to be useful in characterizing single photon
sources \cite{pathak2010recent,verma2008higher} used in quantum cryptography \cite{shukla2014protocols}.
In short, nonclassical features of quantum states are very important
and the same has been studied for various families of quantum states.
One such family of quantum states is called intermediate states \cite{verma2010generalized,pathak2014wigner,verma2008higher}.
These states are interesting because any state of these family can
be reduced to various other quantum states at different limits of
the parameters which define an intermediate state. 

First intermediate state was formally introduced by Stoler et al.,
in 1985 \cite{stoler1985binomial}. The state is referred to as Binomial
state (BS) \cite{stoler1985binomial} and can be defined as 
\begin{equation}
|p,M\rangle=\stackrel[n=0]{M}{\sum}B_{n}^{M}(p)|n\rangle=\stackrel[n=0]{M}{\sum}\left[^{M}C_{n}\,p^{n}(1-p)^{M-n}\right]^{1/2}|n\rangle,\label{eq:binomial}
\end{equation}
where $B_{n}^{M}(p)$ is the probability amplitude
of the binomial state which corresponds to the occurrence of $n$ photons
with equal probability $p$ obtained in $M$ independent ways \cite{stoler1985binomial}.  Mathematically,
the binomial state is equivalent to a molecular system having same
photon emitting probability $p$ from the different energy levels
of the excited states of the molecule which undergoes the M level
vibrational relaxation \cite{fan1999new}. Binomial state being an intermediate
state,  reduces to various existing states at different limits.
For example, it reduces to a (a) vacuum state $|0\rangle$ (if $p\text{\ensuremath{\rightarrow}}0,$
$M\rightarrow0$), (b) number state $|n\rangle$ (if $p\rightarrow1$
and $M$ is finite), (c) coherent state with real amplitude $|\alpha\rangle$
(if $p\rightarrow0$, $M\rightarrow\infty$ with $pM=\alpha^{2}({\rm constant})$). It is interesting to note that coherent states are closest to classical
states and the number states are the most nonclassical states. Thus, fundamentally
different states of electromagnetic field can be obtained as limiting
cases of BS. Naturally, properties of BS has been
studied since long \cite{agarwal_2012}.

The interest on the BS is not restricted to the state of the form
Eq. (\ref{eq:binomial}), it has been extended to various variants
of BS, too. Specifically, in Refs. \cite{agarwal1992negative,barnett1998negative}
negative binomial state was proposed, and subsequently its properties
were studied in Refs. \cite{verma2008higher}. Similarly, reciprocal
binomial state was introduced in Ref. \cite{moussa1998generation}
and studied in \cite{pathak2014wigner,verma2008higher}. Further,
a couple of generalized binomial states (GBS)\footnote{If BS can be obtained as a special (limiting) case of a quantum state,
then that quantum state would be referred to as GBS.} have been proposed \cite{fan1999new} and their nonclassical properties
have also been investigated \cite{verma2008higher,pathak2014wigner}.
More interestingly, possible applications of GBS have been explored
in the field of quantum computation \cite{franco2009quantum}. In
what follows, we aim to study lower- and higher-order nonclassicality
of a particular version of GBS which was introduced by Fan and Liu
\cite{fan1999new} and referred to as a new GBS (NGBS). 

In the introductory work of Fan and Liu, a few nonclassical properties
of this state (e.g., squeezing, and sub-Poissonian photon statistics)
was investigated. These were lower-order nonclassical features, but
no attention was paid to higher-order nonclassical properties of NGBS.
It was natural, as at that time, higher-order nonclssicality was not
of much interest (of course Lee \cite{lee1990higher} and Hong and
Mandel (HM) \cite{hong1985generation}, had already introduced the notion
of higher-order antibunching and higher-order squeezing \cite{giri2017nonclassicality}),
but in the recent past higher-order nonclassicality has been reported
theoretically in \cite{alam2018higher,verma2010generalized} and experimentally
in \cite{allevi2012high,allevi2012measuring}. Further, their applications
(specially applications of higher-order entanglement) have been reported
in establishing quantum supremacy \cite{arrazola2017quantum}. Motivated
by these facts (specially the fact that higher-order nonclassical
properties of NGBS has not yet been studied), and the fact that a
bunch of interesting lower-order nonclassical properties of NGBS is
yet to be investigated, in what follows we study the possibilities
of observing various lower-order and higher-order nonclassical features
for NGBS. Specifically, we report  higher-order antibunching (HOA), higher-order sub-Poissonian photon statistics (HOSPS) and higher-order
squeezing (HOS) of NGBS. We also report nonclassical features present in NGBS using Agarwal Tara criterion, Vogel's criterion and  Wigner
function. These criteria
of nonclassicality are essentially witnesses of nonclassicality as they do not provide any quantitative estimation
of the amount of nonclassicality present in the system. To address
this issue, we have also computed nonclassical volume (volume of the negative
part of Wigner function) which is a measure of the amount of nonclassicality.

Here it would be apt to note that to obtain the nonclassical volume, we
have first derived an analytic expression for the Wigner function
of any finite superposition of Fock states (qudits), and have subsequently
used that to quantify the nonclassicaity present in NGBS. Further,
NGBS may be prepared by generalizing  the exisiting proposals for experimental generation of a family of BS including some GBSs (cf.
 \cite{moussa1998generation,valverde2003generation,franco2006single,franco2010efficient}) and the schemes used for the experiemntal realization of other intermediate
states like photon added coherent state \cite{zavatta2004quantum}. If NGBS can be prepared, then the prepared state is to be characterized
by state tomography. Keeping this possibility in mind, optical tomographs
for NGBS are also produced in this work.

The rest of the paper is organized as follows. In the next section,
we formally introduce NGBS and describe some identities related to the moments of NGBS. In Section
\ref{sec:Nonclassical-properties-of}, we introduce a set of moment-based
criteria of nonclassicality, and establish the existence of nonlcassicality
in NGBS by using those criteria. A quantitative measure of nonclassicality
is also provided  in the form of nonclassical volume of Wigner function. In Section \ref{subsec:Optical-tomogram}, we provide
optical tomograms for NGBS. Finally, the paper is concluded in Section
\ref{sec:Conclusion}.

\section{Generalized binomial state of our interest and analytic expressions of moments }

The NGBS introduced by Fan and Liu \cite{fan1999new} is defined as 

\begin{equation}
|M,p,q\rangle=\stackrel[n=0]{M}{\sum}B_{n}^{M}(p,\,q)|n\rangle,\label{eq:NGBS}
\end{equation}
where

\begin{equation}
B_{n}^{M}(p,\,q)=\left[\frac{p}{1+Mq}{}^{M}C_{n}\left(\frac{p+nq}{1+Mq}\right)^{n-1}\left(1-\frac{p+nq}{1+Mq}\right)^{M-n}\right]^{1/2},\label{eq:NGBS-coefficient}
\end{equation}
with $n=0,1,2,3,....M$ and  $0<p<1$. Mathematical expression of NGBS was given by
Fan and Liu by using the Abel\textquoteright s generalization of the
binomial formula, i.e., $p$ is replaced by $\frac{p+nq}{1+Mq}$ in
Eq. (\ref{eq:binomial}), where the value of $q$ may be positive,
zero or negative but in order to satisfy the condition for the probability
amplitude, $B_{n}^{M}(p,\,q)>0$ , $q$ must obey $q\geq Max\left\{ -\frac{p}{M},-\frac{(1-p)}{M}\right\} $.
From Eq. (\ref{eq:NGBS-coefficient}) it is clear that when $q=0$,
NGBS (\ref{eq:NGBS}) reduces to BS (\ref{eq:binomial}). However, it
is interesting to investigate the change of nonclassical properties
for other values of $q$. Physically, the objective behind the generalization involved here was that the photon emitting probability $p$ of a molecular system is
different for all the energy levels of an excited state during a nonlinear
process \cite{fan1999new}. This is why the factor $q$ was introduced to model the  generalized process. It
is also noted that for large values of $M$, $q$ is small for the
fixed values of $p$. Therefore, from the experimental point of view
the NGBS appears to be more general and more realistic state.

There exists a large number of  nonclassical criteria. Many of them are  based on the moments of creation and annihilation operators
($a^{\dagger}$and $a$). This section is focused on getting a general
expression of moments  of annihilation and creation operators for NGBS. 
Repeatedly applying the annihilation operator on NGBS we obtain

\begin{equation}
\begin{array}{lcc}
a|M,p,q\rangle & = & \left[\frac{pM}{(1+Mq)^{M}}\stackrel[n=0]{M}{\sum}\frac{(M-1)!}{(n-1)![(M-1)-(n-1)]!}\left(p+nq\right){}^{n-1}\left(1-p+\left(M-n\right)q\right){}^{M-n}\right]^{1/2}|n-1\rangle,\end{array}\label{eq:calculation 1}
\end{equation}

\begin{equation}
\begin{array}{lcc}
a^{2}|M,p,q\rangle & = & \left[\frac{pM(M-1)}{(1+Mq)^{M}}\stackrel[n=0]{M}{\sum}\frac{(M-2)!}{(n-2)![(M-2)-(n-2)]!}\left(p+nq\right){}^{n-1}\left(1-p+\left(M-n\right)q\right){}^{M-n}\right]^{1/2}|n-2\rangle,\end{array}\label{eq:calculation 2}
\end{equation}
and

\begin{equation}
\begin{array}{lcc}
a^{l}|M,p,q\rangle & = & \left[\frac{M!}{(M-l)!}\stackrel[n=0]{M}{\sum}\,^{M-l}C_{n-l}\,\frac{p}{1+Mq}\left(\frac{p+nq}{1+Mq}\right){}^{n-1}\left(1-\frac{p+nq}{1+Mq}\right){}^{M-n}\right]^{1/2}|n-l\rangle.\end{array}\label{eq:calculation 3}
\end{equation}
Eq. (\ref{eq:calculation 3}) implies

\begin{equation}
\begin{array}{lcc}
\langle M,p,q|a^{\dagger k} & = & \langle n'-k|\left[\frac{M!}{(M-k)!}\stackrel[n'=0]{M}{\sum}\,^{M-k}C_{n'-k}\,\frac{p}{(1+Mq)}\left(\frac{p+n'q}{1+Mq}\right){}^{n'-1}\left(1-\frac{p+n'q}{1+Mq}\right){}^{M-n'}\right]^{1/2}.\end{array}\label{eq:equation 4}
\end{equation}
Therefore,

\begin{equation}
\begin{array}{lcc}
\langle a^{\dagger k}a^{l}\rangle & = & \frac{pM!}{(1+Mq)}\left[\stackrel[n=0]{M}{\sum}\frac{1}{(n-l)!}\sqrt{\frac{1}{(M-n)!\left(M-n+l-k\right)!}\left(\frac{p+nq}{1+Mq}\right){}^{n-1}\left(\frac{p+\left(n-l+k\right)q}{1+Mq}\right){}^{n-l+k-1}}\right.\\
 & \times & \left.\sqrt{\left(1-\frac{p+nq}{1+Mq}\right){}^{M-n}\left(1-\frac{p+\left(n-l+k\right)q}{1+Mq}\right){}^{M-n+l-k}}\right].
\end{array}\label{eq:akal}
\end{equation}
In what follows, we will see that the above analytic expressions will
essentially lead to analytic expressions for various witnesses of
nonclassicality. 

\section{Nonclassical properties of NGBS \label{sec:Nonclassical-properties-of}}

\subsection{Higher-order antibunching}

\begin{figure}
\centering{}
\subfigure[]{\includegraphics[scale=0.6]{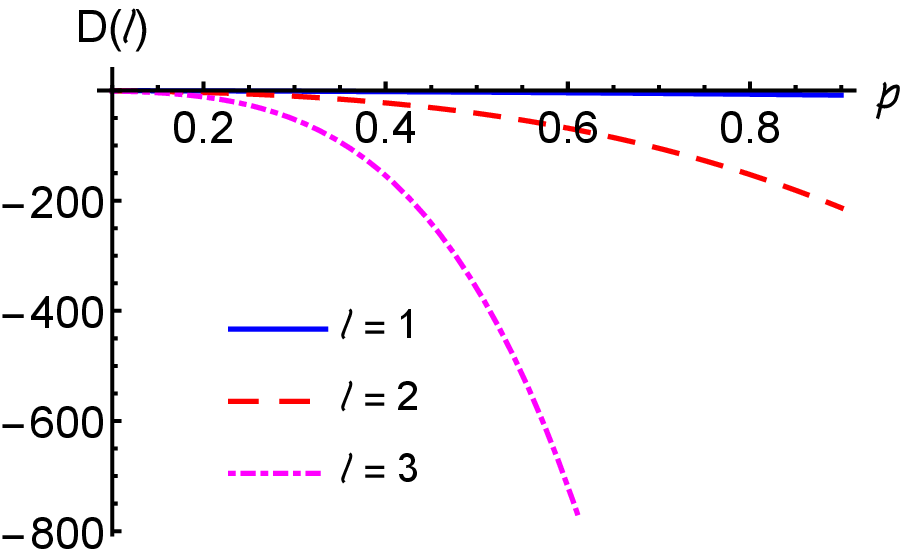}} \quad\subfigure[]{ \includegraphics[scale=0.6]{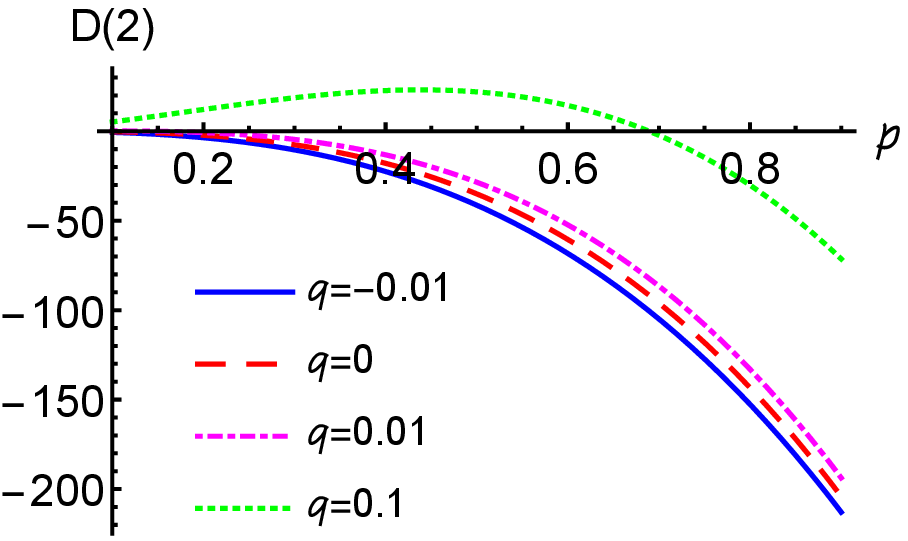}}\quad \subfigure[]{\includegraphics[scale=0.6]{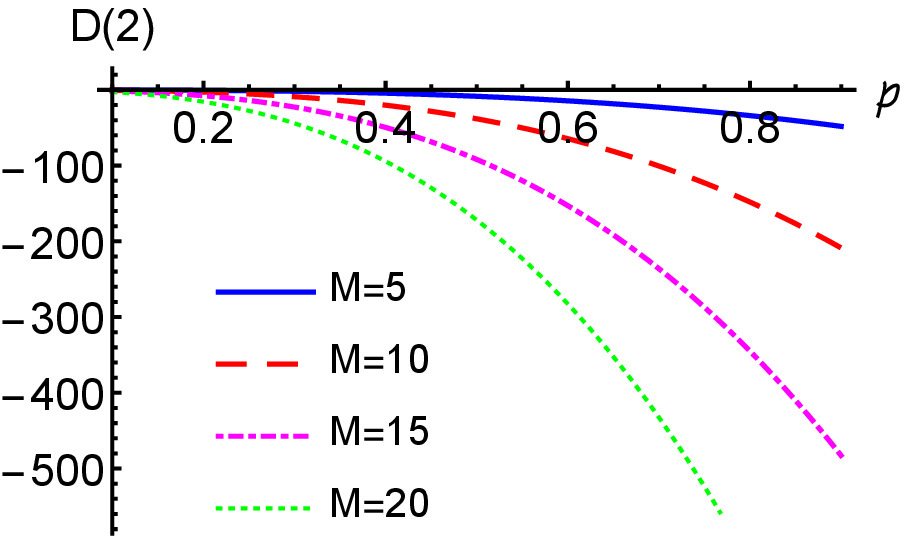}}
\caption{\label{fig:HOA}(Color online) Variation of HOA for NGBS is shown here with probability
$p$ for the fixed values of (a) $M=10$ and $q=-0.02$ with different
order number, (b) $M=10$, and $l=2$ with different values of $q$,
and (c) $l=2$ and $q=-0.005$ with different values of $M$.}
\end{figure}

The phenomena of antibunching is closely related to the photon
statistics of a state. This phenomenon corresponds to a physical situation involving two photons (or 2 modes), in which the  probability of getting two photons simultaneously is
less than the probability of getting them separately (one-by-one). Generalization of this idea into multi-photon regime leads to the notion of HOA. In fact, in 1990, Lee \cite{lee1990higher}
introduced the concept of the HOA using the theory of majorization.
Lee's criterion for HOA was subsequently modified by Ba An \cite{an2002multimode}
and Pathak and Garcia, \cite{pathak2006control}. As per the criterion of Pathak and Garcia, a quantum state is considered to be higher-order antibunched if it satisfies the following inequality \cite{pathak2006control} 

\begin{equation}
D(l)=\langle N^{(l+1)}\rangle-\langle N\rangle^{l+1}=\langle a^{\dagger l+1}a^{l+1}\rangle-\langle a^{\dagger}a\rangle^{l+1}<0,\label{eq:antibunching}
\end{equation}
where $N=a^{\dagger}a$ is the number operator and $N^{(l+1)}=a^{\dagger l+1}a^{l+1}$
is the $l$th order factorial moment, respectively. Eq. (\ref{eq:antibunching})
corresponds to the $l$ th order antibunching criterion. For $l=1$, it reduces
to the lower-order (conventional) antibunching criterion and for $l\geq 2$ it corresponds to higher-order
antibunching criterion. In this article, we have investigated HOA using
the criterion (\ref{eq:antibunching}). We have clearly observed the
existence of HOA in NGBS (cf. Figs. \ref{fig:HOA}(a)-(c)). The negative
part of the curves ensures that NGBS  satisfies the inequality  (\ref{eq:antibunching})
and hence NGBS is higher-order antibunched. In Figs. \ref{fig:HOA}(a) and
(c), we observe that the depth of the HOA witness increases with the increase
of order number $l$ and the dimension $M$, 
whereas in  Fig. \ref{fig:HOA}(b), it decreases with $q$ and for the large values of $q$ it
becomes positive, i.e., the signature of HOA is found to be lost for large values of $q$.

\subsection{Higher-order sub-Poissonian photon statistics}

\begin{figure}
\centering{}

\subfigure[]{\includegraphics[scale=0.6]{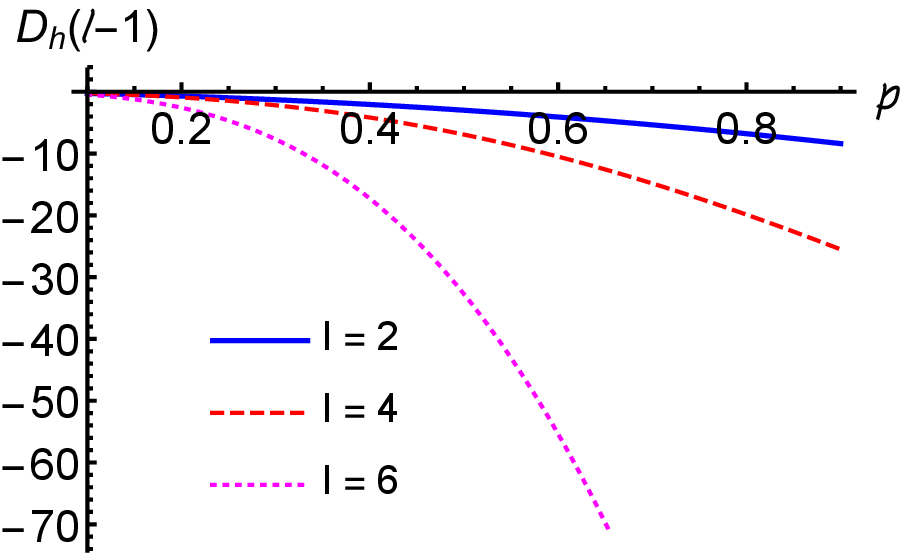}} \subfigure[]{\includegraphics[scale=0.6]{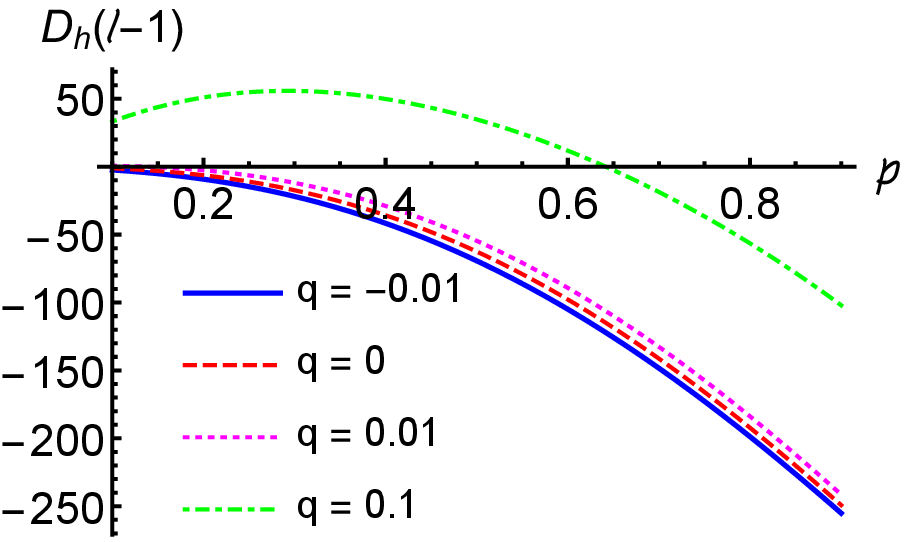}} \subfigure[]{ \includegraphics[scale=0.6]{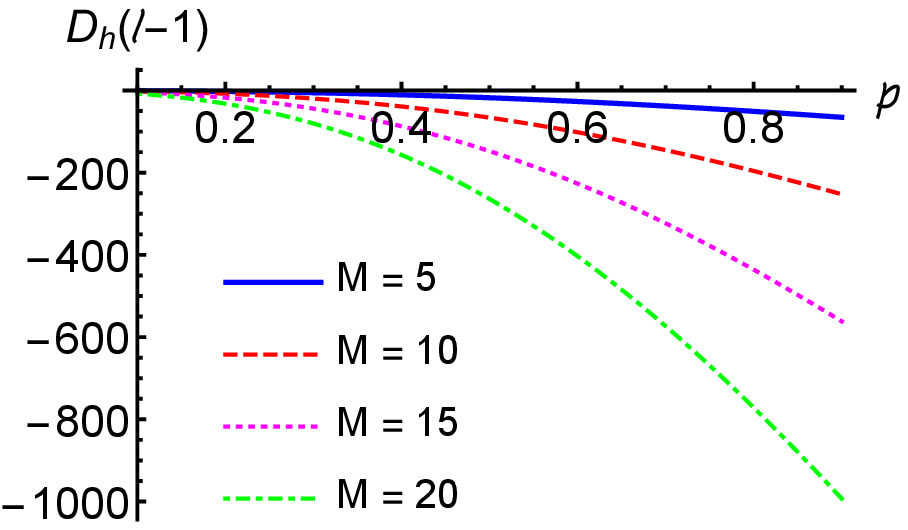}}

\caption{\label{fig:HOSPS}(Color online) Variation of HOSPS criterion for NGBS is shown here
with probability $p$ for the fixed values of (a) $M=10$ and $q=-0.02$
with different order number, (b) $M=10$, and $l=2$ with different
values of $q$, and (c) $l=2$ and $q=-0.005$ with different values
of $M$. }
\end{figure}

Higher-order nonclassical feature associated with the photon statistics
of a quantum state of radiation field is usually studied through the
witness of HOSPS. Here it may be noted that HOSPS is the higher-order
analogue of the frequently investigated sub-Poissonian photon statistics, and 
HOSPS is observed for a state if higher-order moment of the photon number
for that state is found to be less than the corresponding moment for a  Poissonian state, i.e., $\langle\left(\Delta N\right)^{l}\rangle<\langle\left(\Delta N\right)^{l}\rangle|_{Poissonian}$.
The generalized moment-based
criterion to observe HOSPS is given as \cite{verma2010generalized}

\begin{equation}
D_{h}(l-1)=\sum_{r=0}^{l}\sum_{k=0}^{r}S_{2}(r,k)\,^{l}C_{r}(-1)^{r}D(k-1)\langle N\rangle^{l-r}<0\label{eq:HOSPS}
\end{equation}
where $S_{2}(r,k)$ is the Stirling number of the second kind. The inequality
in Eq. (\ref{eq:HOSPS}) is the condition for the $(l-1)$th order
nonclassicality, and for $l\geq3$ it leads to the condition for HOSPS.
We computed analytic expression for $D_{h}(l-1)$
using Eqs. (\ref{eq:akal}) and (\ref{eq:antibunching}), and the
corresponding results are illustrated in Figs. \ref{fig:HOSPS}(a)-(c), where
the negative parts in figures depict the existence of HOSPS in NGBS. In Figs. \ref{fig:HOSPS}(a)
and (c), we can also observe that the depth of the witness of HOSPS increases
with order number $l$ and dimension $M$, 
but it decreases with parameter $q$ (cf. Fig. \ref{fig:HOSPS}(b)).
From Fig. \ref{fig:HOSPS}(b), it is observed that for large values
of  $q$, HOSPS criterion (\ref{eq:HOSPS}) is not satisfied when
the probability remains below a certain value. However, it is observed
when the probability is greater than that  value. 

\subsection{Higher-order squeezing}

The squeezing phenomenon originates from the Heisenberg uncertainty
relation. The product of fluctuation of two non-commuting operators
in Heisenberg uncertainty relation (uncertainty product) has a minimum
value. For a coherent state the quadrature variances are equal, and their product possesses minimum value. If the
variance of one of the quadrature goes below this  value (in order
to respect the Heisenberg uncertainty relation the other quadrature
must be greater than this  value), the corresponding quadrature is considered to be
 squeezed. The higher-order counterpart of the squeezing is higher-order
squeezing. There are types of  higher-order
squeezing, which are frequently used-  Hong-Mandel squeezing \cite{hong1985generation} and
Hillery type squeezing or amplitude powered squeezing \cite{hillery1987amplitude}.
 To begin with, we investigate the possibility of observing Hong-Mandel squeezing which can be
described by the following criterion

\begin{figure}
\centering{}

\subfigure[]{\includegraphics[scale=0.7]{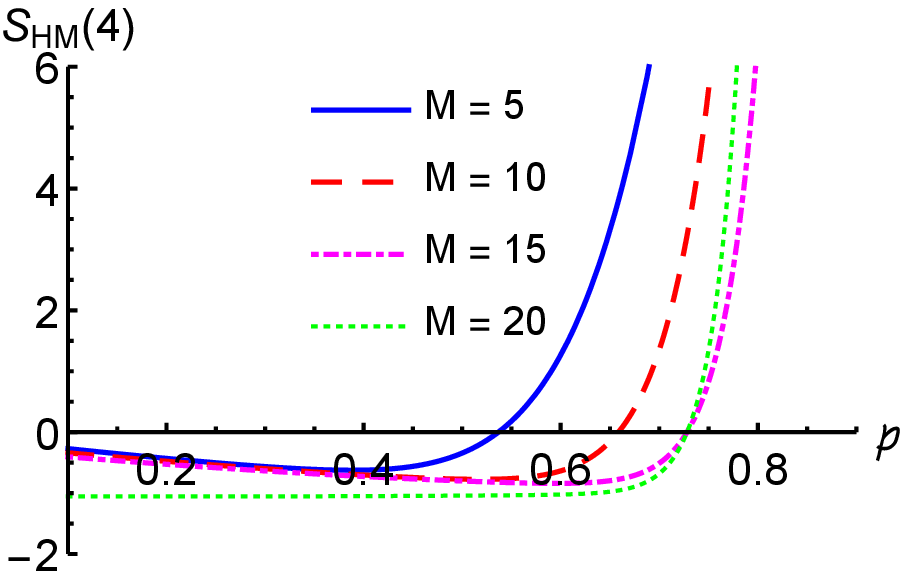}}\quad \subfigure[]{\includegraphics[scale=0.7]{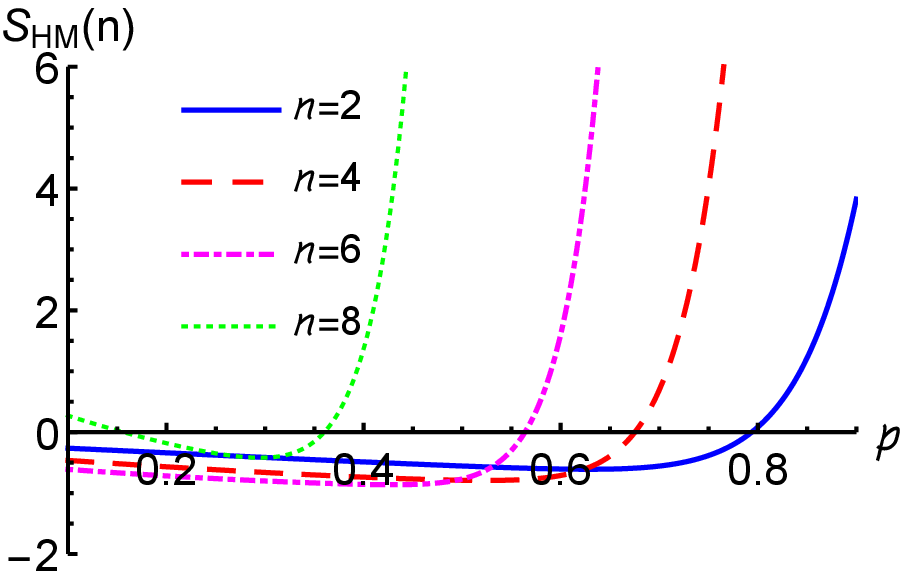}}\\
\subfigure[]{\includegraphics[scale=0.7]{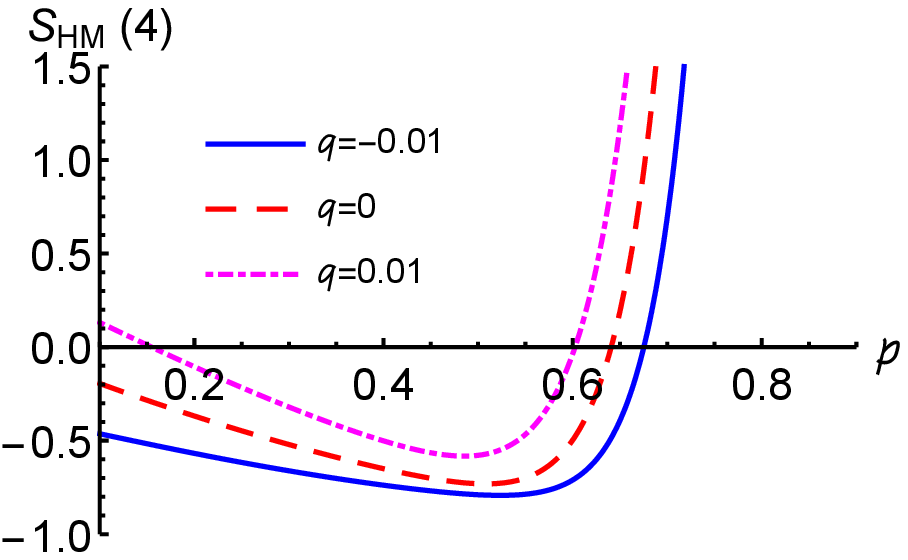}}\quad \subfigure[]{\includegraphics[scale=0.7]{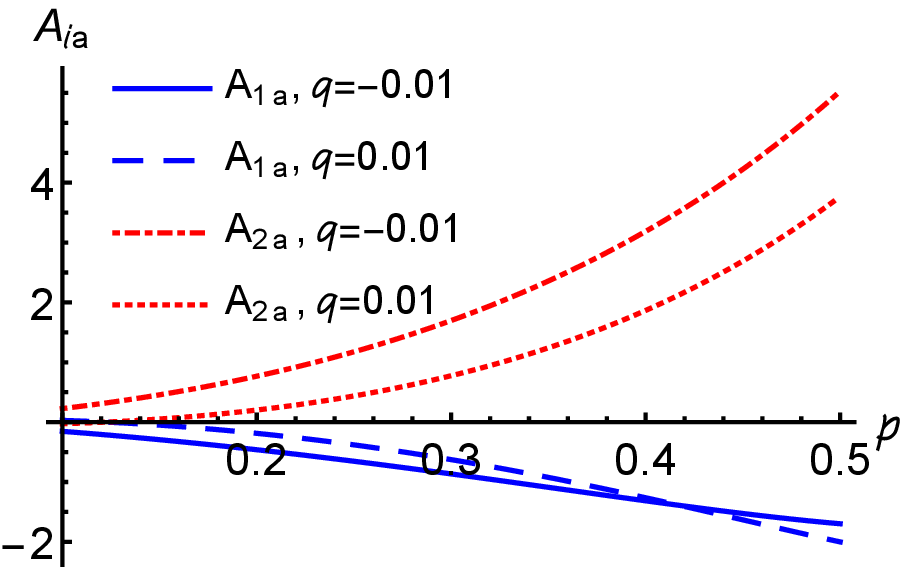}}   

\caption{\label{fig.HOS}(Color online) HOS of Hong-Mandel type for NGBS for (a) $q=-0.005$
with different value of $M$, (b) $M=10$ and $q=-0.005$ with different
values of $n$ and (c) $M=10$ with different values of $q$. (d)
Higher-order squeezing Hillery type for $q=-0.01$ and $M=10$.}
\end{figure}

\begin{equation}
S_{HM}(n)=\frac{\langle(\Delta X)^{n}\rangle-\left(\frac{1}{2}\right)_{\frac{n}{2}}}{\left(\frac{1}{2}\right)_{\frac{n}{2}}}<0,\label{eq:HM-criterion}
\end{equation}
where

\begin{equation}
\langle(\Delta X)^{n}\rangle=\sum_{r=0}^{n}\sum_{i=0}^{\frac{r}{2}}\sum_{k=0}^{r-2i}(-1)^{r}\frac{1}{2^{\frac{n}{2}}}(2i-1)!!\,^{r-2i}C_{k}\,^{n}C_{r}\,^{r}C_{2i}\langle a^{\dagger}+a\rangle^{n-r}\langle a^{\dagger k}a^{r-2i-k}\rangle<\left(\frac{1}{2}\right)_{\frac{n}{2}}=\frac{1}{2^{\frac{n}{2}}}(n-1)!!,\label{eq:Hong-mandel}
\end{equation}
and the symbol $\left(x\right)_{n}$ is the conventional Pochhammer
symbol. The inequality in Eq. (\ref{eq:HM-criterion}) is investigated
analytically using Eqs. (\ref{eq:akal}) and (\ref{eq:HM-criterion})
and the corresponding results are depicted in Figs. \ref{fig.HOS}(a)-(c) where it is observed that depth of HOS(HM) increases with M, decreases with q and region of nonclassicality decreases with increasing order number. 

Hillery treated the HOS in a different way. Instead of higher-order
moment, he introduced amplitude powered quadratures. The variance of
this quadrature is used to investigate HOS. The criterion to obtain
Hillery's amplitude powered squeezing is described as \cite{alam2017lower}

\begin{equation}
A_{i,a}=(\Delta Y_{i,a})^{2}-\frac{1}{2}|\langle\left[Y_{1,a},Y_{2,a}\right]\rangle|<0,\label{eq:Hillery3}
\end{equation}
where $Y_{1,a}=\frac{a^{l}+a^{\dagger l}}{2}$ and $Y_{2,a}=\frac{-i(a^{l}+a^{\dagger l})}{2}$
are the amplitude powered quadrature. In this article, we have calculated
HOS with amplitude squared squeezing i.e., for $l=2$, using Hillery's
HOS criterion (\ref{eq:Hillery3}) and Eq. (\ref{eq:akal}). The result
is exhibited in Fig. \ref{fig.HOS}(d) where the negative values of
the  variances clearly show the existence of HOS of Hillery type in NGBS. 

%

\subsection{Agarwal Tara criterion}

In 1992, Agarwal and Tara \cite{agarwal1992nonclassical} introduced
a nice criterion to investigate nonclassical phenomenon based on the
appropriate generalization of Mandel's Q parameter. This criterion may
reveal nonclassicality even when the criteria for the determination of squeezing and sub-Poissonian photon statistics fail to detect any signature of nonclassicality. Specifically, they introduced
a criterion for nonclassicality involving the matrix of normally ordered
moments $m_{l}=\langle a^{\dagger l}a^{l}\rangle$ and moments of
number distribution $\mu_{l}=\langle(a^{\dagger}a)^{l}\rangle$ as follows
\begin{equation}
A_{n}=\frac{\det m^{(n)}}{\det\mu^{(n)}-\det m^{(n)}}<0,\label{eq:Agarwal Tara}
\end{equation}
where $\det m^{(n)}$ and $\det\mu^{(n)}$ are the determinant of
matrix of normally order moments $m_{n}$ and moments
of number distribution $\mu_{n}$, respectively. The matrices $m^{(n)}$
and $\mu^{(n)}$ can be written as 
\begin{center}
$m^{(n)}=\left[\begin{array}{cccc}
1 & m_{1} & \cdots & m_{n-1}\\
m_{1} & m_{2} & \cdots & \vdots\\
\vdots & \vdots & \ddots & \vdots\\
m_{n-1} & \cdots & \cdots & m_{2n-2}
\end{array}\right]$ and $\mu^{(n)}=\left[\begin{array}{cccc}
1 & \mu_{1} & \cdots & \mu_{n-1}\\
\mu_{1} & \mu_{2} & \cdots & \vdots\\
\vdots & \vdots & \ddots & \vdots\\
\mu_{n-1} & \cdots & \cdots & \mu_{2n-2}
\end{array}\right]$.
\par\end{center}

\begin{figure}
\centering{}
\subfigure[]{\includegraphics[scale=0.7]{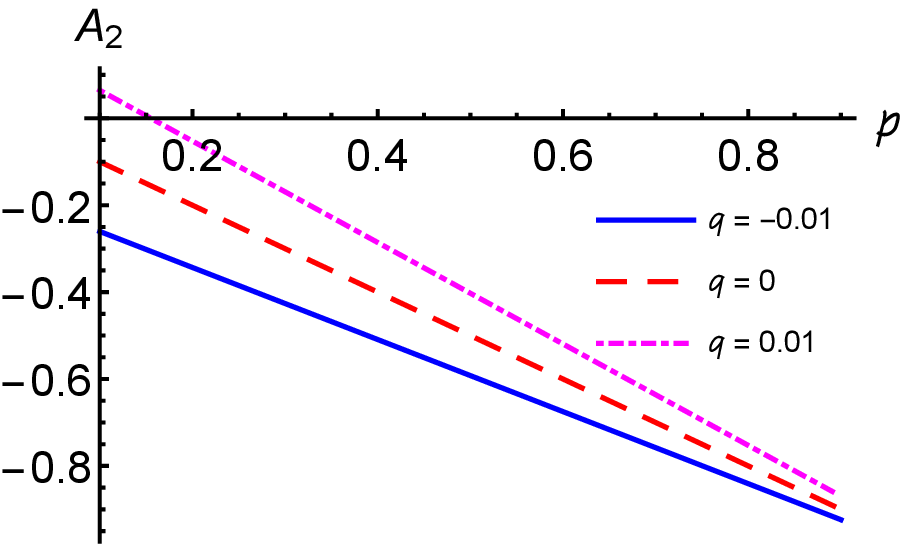}}\quad  \subfigure[]{\includegraphics[scale=0.7]{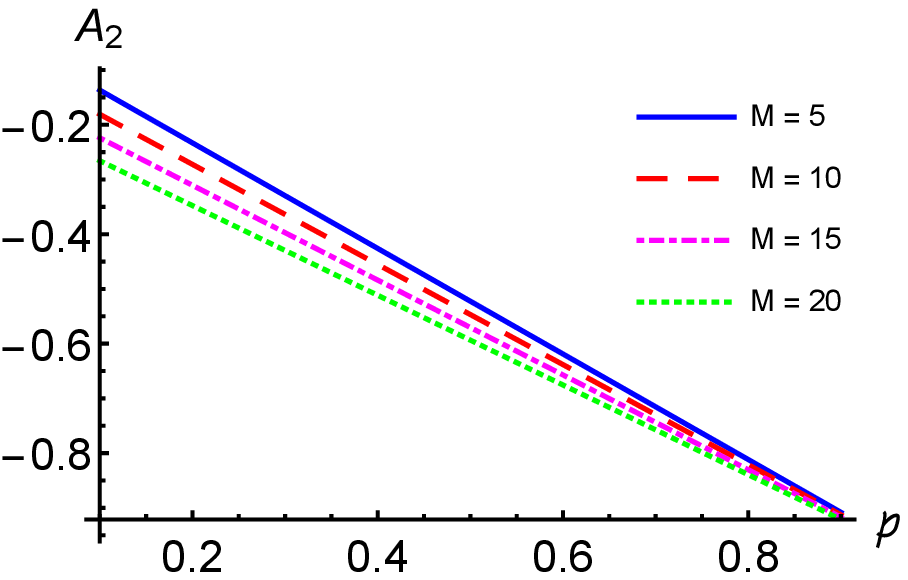}}\\
\subfigure[]{\includegraphics[scale=0.7]{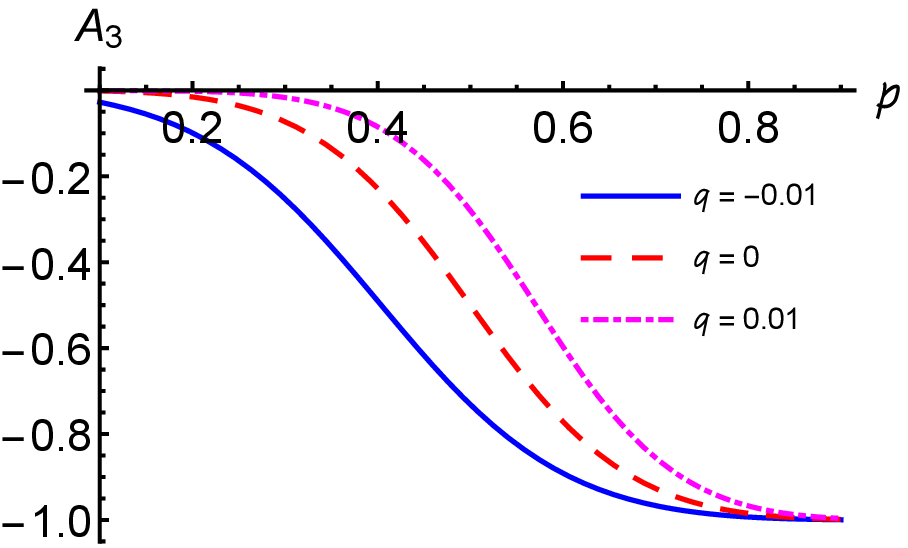}}\quad \subfigure[]{\includegraphics[scale=0.7]{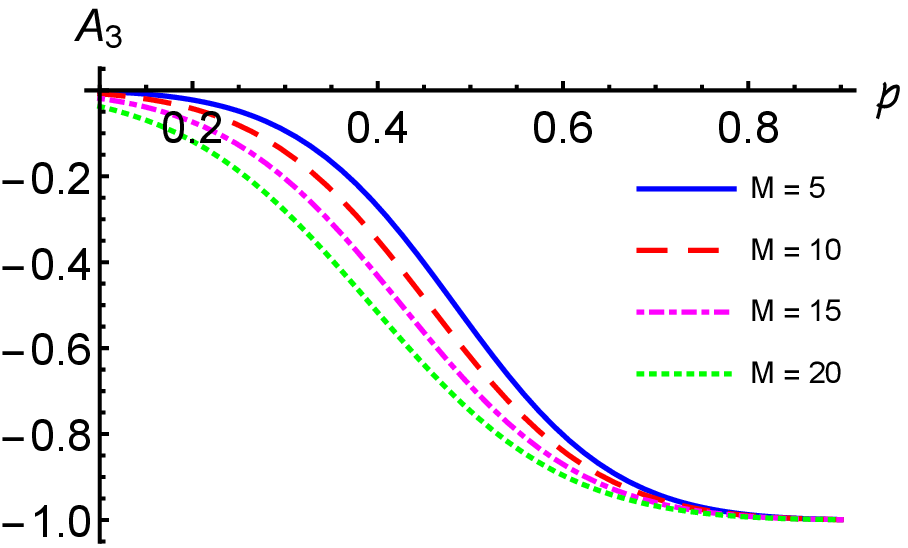}}

\caption{\label{fig:Agarwal-Tara}(Color online) Variation of Agarwal Tara nonclassicality
criterion with respect to probability $p$. Variation of $A_{2}$
for (a) $M=10$ and different values of $q$, (b) for $q=-0.005$ and different values of $M$. Similarly, Variation of
$A_{3}$ for (c) $M=10$ and different values of $q$, (d) for $q=-0.005$ and different values of $M$. }
\end{figure}

The nonclassical phenomenon is observed for  $-1\leq A_{n}\leq0$.
Boundary values $A_{n}=0$ and $A_{n}=-1$ correspond to the closest-to-classical (coherent or its mixture) state and Fock  (most nonclassical) state, respectively.
In brief, for a nonclassical state, $A_{n}$ is always negative and bounded
by the value $-1$ when the state become maximally nonclassical.
We have computed  analytical expression of $A_{n}$ for particular values of $n$ using Eq. (\ref{eq:akal})
and have investigated the variation of  nonclassical phenomenon witnessed through Agarwal-Tara criterion. The results are shown in 
Figs. \ref{fig:Agarwal-Tara}(a)-(d) where the negative regions of
the curves depict nonclassicality. It is observed that negative values
of $A_{n}$ increase with probability $p$ and reaches $-1$ when $p$ approaches
unity, indicating that for $p=1$  NGBS is maximally nonclassical.
From Figs. \ref{fig:Agarwal-Tara}(a) and (c), we have seen that for
both $A_{2}$ and $A_{3}$ the depth of $A_{n}$ in the negative region  decreases
with increase of parameter $q$. Similarly, from Figs. \ref{fig:Agarwal-Tara}(b)
and (d), we have also observed that the depth of nonclassicality as witnessed through $A_{n}$  increases with
 $M$. This is consistent with the nonclassical features observed through the other criteria.

\subsection{Vogel's criterion}

\begin{figure}
\centering{}
\subfigure[]{\includegraphics[scale=0.7]{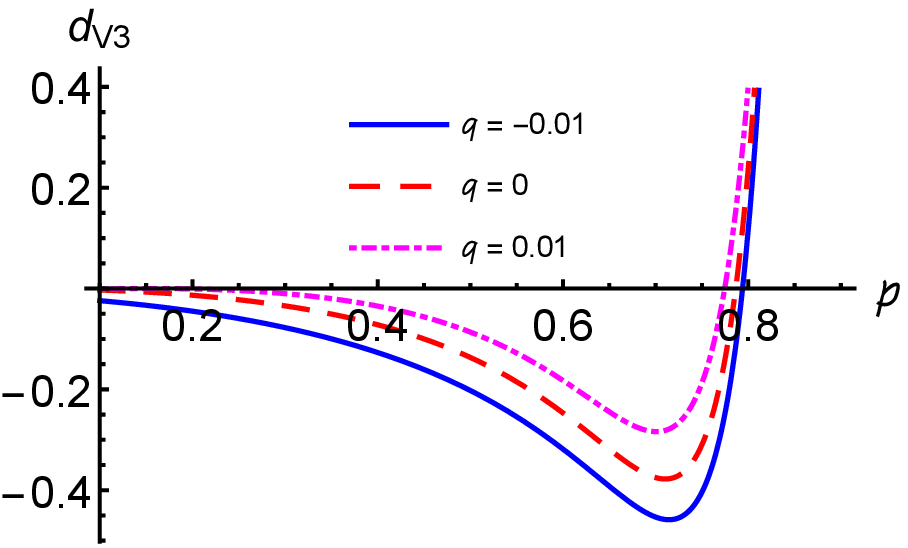}}\quad 
\subfigure[]{\includegraphics[scale=0.7]{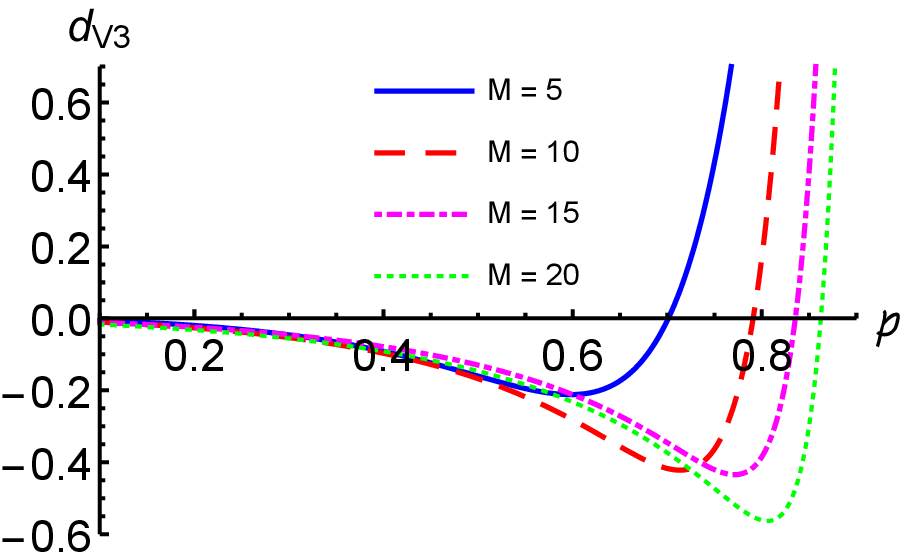}}\\
\subfigure[]{\includegraphics[scale=0.7]{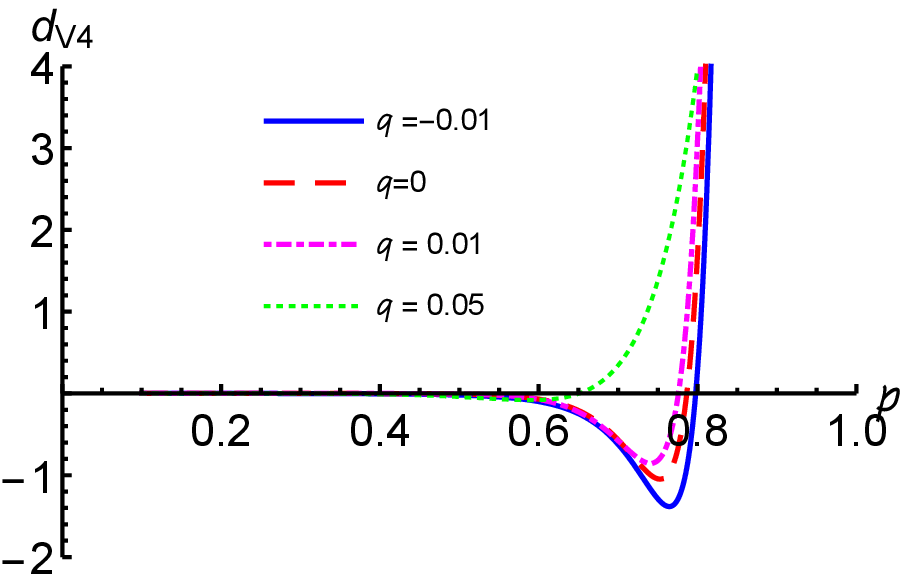}}\quad
\subfigure[]{\includegraphics[scale=0.7]{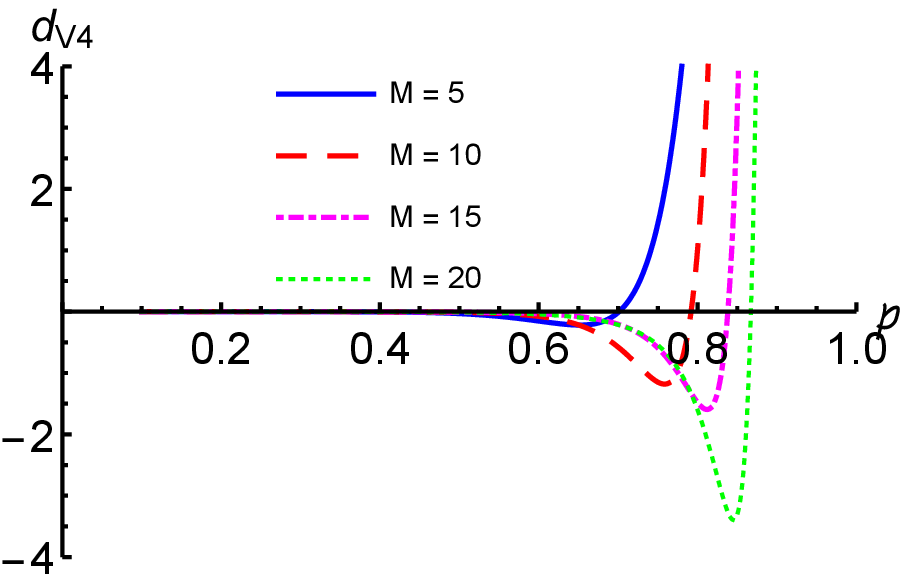}}

\caption{\label{fig.vogel} (Color online) Variation of $d_{v3}$ with respect to probability
$p$ for (a) $M=10$ and different values of $q$, (b) $q=-0.005$
and different values of $M$. Similarly, variation of $d_{v4}$ with
respect to probability $p$ for (c) $M=10$ and different values of $q$ and (d) $q=-0.005$ and different values
of $M$.}
\end{figure}

In Ref. \cite{shchukin2005nonclassical}, Vogel et
al., derived a criterion for nonclassicality in terms of moments of
$a^{\dagger}$ and $a$ operators in the form of $N\times N$ matrix.  The determinant of the $N\times N$ matrix is given by

\begin{equation}
d_{vN}=\left|\begin{array}{ccccc}
1 & \langle a\rangle & \langle a^{\dagger}\rangle & \langle a^{2}\rangle & .....\\
\langle a^{\dagger}\rangle & \langle a^{\dagger}a\rangle & \langle a^{\dagger2}\rangle & \langle a^{\dagger}a^{2}\rangle & .....\\
\langle a\rangle & \langle a^{2}\rangle & \langle a^{\dagger}a\rangle & \langle a^{3}\rangle & .....\\
\langle a^{\dagger2}\rangle & \langle a^{\dagger2}a\rangle & \langle a^{\dagger3}\rangle & \langle a^{\dagger2}a^{2}\rangle & .....\\
\vdots & \vdots & \vdots & \vdots & \ddots
\end{array}\right|_{N\times N}.\label{eq:vogel-criterion}
\end{equation}
For a nonclassical state  at least
one of  the determinants  $d_{vN}$ of the matrix
(\ref{eq:vogel-criterion}), should be negative  i.e., 

\begin{equation}
d_{vN}<0,\,\,\,\,\rm{for} \,\,N=3,\,4, .... \,.
\end{equation}
For  $N=2$, $d_{vN}$ is positive because in this case  $d_{v2}$ represents the incoherent part of the photon number. The negativity of any such sub determinant is the sufficient condition for nonlclassicality. It is also reported that all such moments of $a^{\dagger}$and $a$ can be realized using beam spliters and homodyne correlation measurements. Here we have reported the result for $N=3$ and $4$ in Figs. \ref{fig.vogel}(a)-(d). The plots clearly illustrate the fact that NGBS is highly nonclassical. Further, in consistent with other criteria discussed here,  it also shows that with the increase of parameter $q$, the depth and region of nonclassicality witness decrease  (cf. Figs. \ref{fig.vogel}(a) and (c)). But this type of  consistency is not observed with  the increase of dimension $M$. In this case the depth of the nonclassicality witness increases with the increase of dimension $M$ only when the probability is high  (cf. Figs. \ref{fig.vogel}(b) and (d)).  This is a very general criterion and many of the above mentioned criteria (including Agarwal-Tara criterion and Hillery's criterion of HOS) can be obtained as special cases of this criterion.  

\subsection{Wigner function }

In this section, we shall investigate Wigner qasiprobability distribution
function for NGBS. The NGBS can be written as a finite dimensional
Fock superposition state (FSS) and can be viewed as a qudit. Any such Fock superposition state
has a general form 
\begin{equation}
|\psi\rangle=\sum_{n=0}^{N}c_{n}|n\rangle,\label{eq:quantum state1}
.\end{equation}
Such a state has to be nonclassical as we can imagine that this finite dimensional FSS  has infinitely many holes in its photon
number distribution as for this state $P(i:i>N)=0,$ where $P(i)=|c_{i}|^{2}$
is the probability of finding  $i$th quantum
state $i\rangle$ which has $i$ photon\footnote{Any pure quantum state can be expressed as superposition of Fock
states. If in photon number distribution (PND) we find $P_{n}=|c_{n}|^2=0$ for any $n$, we say that a hole is created at that position in the photon number distribution.}
state having holes in their PND is nonclassical
in nature. For example, consider an arbitrary state $\rho=\int P(\alpha)|\alpha\rangle\langle\alpha|d^{2}\alpha$
having photon number distribution $P_{n}=\int P(\alpha)|\langle n|\alpha\rangle|^{2}d^{2}\alpha$. 
Since the values $\langle n|\alpha\rangle|^{2}>0$ , therefore, $P_{n}\ne0$
when $P(\alpha)$ is a true probability density.

NGBS is just a special form of this state (\ref{eq:quantum state1})
where $c_{n}$ should be replaced by Eq. (\ref{eq:NGBS-coefficient}).
In what follows we provide analytic expressions for Wigner function for finite dimensional FSS of  the form (\ref{eq:quantum state1}), in general, and subsequently
to witness and quantify nonclassicality present in NGBS by plotting
the analytic expression.

In fact, Wigner function and other quasi-probability distributions
have been studied for a long time. Naturally, efforts had been made to construct
analytic expressions of Wigner function of the states of the form (\ref{eq:quantum state1}).
To be precise, in \cite{moya1993series} the authors commented, ``The
Wigner function is usually expressed in an integral form which is
not always easy to compute.'' Keeping that in mind they provided
(cf. Eq. (16) of \cite{moya1993series} ) the following analytic expression
of Wigner function: 
\begin{equation}
W(\alpha)=\frac{2}{\pi}\sum_{k=0}^{\infty}(-1)^{k}\langle\alpha,k|\rho|\alpha,k\rangle,\label{eq:series wigner}
\end{equation}
where $\rho$ is the density matrix of the state and $|\alpha,k\rangle$
are the displaced number states. Since $|\psi\rangle$ in (\ref{eq:quantum state1})
is a pure state so we have $\rho=|\psi\rangle\langle\psi|$ and consequently,
\[
\begin{array}{lcl}
W(\alpha) & = & \frac{2}{\pi}\sum_{k=0}^{\infty}(-1)^{k}|\langle\alpha,k|\psi\rangle|^{2}\\
 & = & \frac{2}{\pi}\sum_{k=0}^{\infty}(-1)^{k}|\sum_{n}c_{n}\langle\alpha,k|n\rangle|^{2}\\
 & = & \frac{2}{\pi}\sum_{k=0}^{\infty}(-1)^{k}|\sum_{n}c_{n}\langle k|D(\alpha)|n\rangle|^{2}\\
 & = & \frac{2}{\pi}\sum_{k=0}^{\infty}(-1)^{k}|\sum_{n}c_{n}\chi_{kn}(\alpha)|^{2},
\end{array}
\]
 where $D(\alpha)$ is the displacement operator and $\chi_{kn}(\alpha)=\langle k|D(\alpha)|n\rangle$
which can be expressed in compact form as (cf. Eq. 4.6 of \cite{fu1997hypergeometric})
\[
\chi_{nk}(\alpha)=\left\{ \begin{array}{c}
\sqrt{\frac{k!}{n!}}\exp\left(-\frac{|\alpha|^{2}}{2}\right)\alpha^{n-k}L_{k}^{n-k}\left(|\alpha|^{2}\right)\,{\rm if\:}n\geq k\\
\sqrt{\frac{n!}{k!}}\exp\left(-\frac{|\alpha|^{2}}{2}\right)\left(\alpha^{*}\right)^{k-n}L_{n}^{k-n}\left(|\alpha|^{2}\right)\,\,{\rm if\:}n\leq k.
\end{array}\right.
\]
 Till now, the above mentioned series expansion of Wigner function was used
to investigate the quasi-distribution of FSSs. Here, we would like
to note that in contrary to the comment of Cessa and Night in \cite{moya1993series},
it is often difficult to handle the infinite sum present in the series
form of Wigner function. Keeping that in mind, we have derived a compact
form of Wigner function of FSS below in integral form and have shown
that it is free from the trouble of handling infinity. A detail derivation
of the Wigner function is provided in Appendix A. Here we just note the final expression of Wigner function for FSS of above form can be obtained as (\ref{Apprndix-wigner})

\begin{equation}
W(x,p')=\frac{1}{\pi^{\frac{1}{2}}}\sum_{n,n^{\prime}=0}^{N}c_{n}^{*}c_{n^{\prime}}b_{n}^{*}b_{n^{\prime}}{\rm e}^{-\left(x^{2}+p'^{2}\right)}(-1)^{n^{\prime}}2^{n^{\prime}}n!(ip'-x)^{n^{\prime}-n}L_{n}^{n^{\prime}-n}\left(-2y(ip'-x)\right),\,\,\,\,\,n\leq n^{\prime}.\label{eq:wigner final}
\end{equation}

\begin{figure}
\centering{}
\subfigure[]{\includegraphics[scale=0.6]{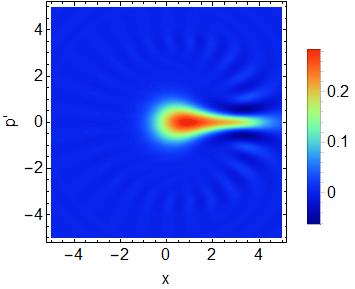}}\quad 
\subfigure[]{\includegraphics[scale=0.6]{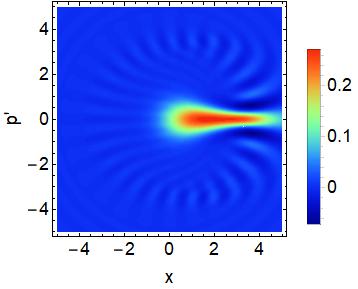}}\\
\subfigure[]{\includegraphics[scale=0.6]{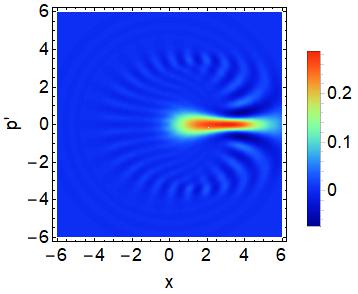}}\quad \subfigure[]{\includegraphics[scale=0.6]{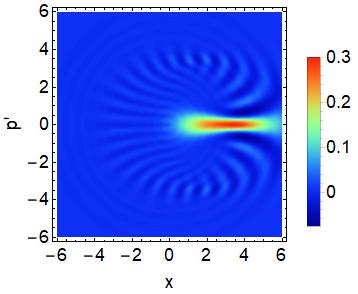}}

\caption{\label{fig:Wigner-GBS}(Color online) Wigner function for the NGBS for (a) $p=0.2$,
(b) $p=0.4$, (c) $p=0.6$ and (d) $p=0.8$, respectively. In all
figures $M=25$ and $q=0.5$.}
\end{figure}

Here we have used $p'$ instead of $p$ because the symbol $p$ is already used for probability.  Now we can use the formalism presented here to obtain Wigner function
for any FSS. However, this paper is focused on NGBS and we restrict ourselves to the construction of Wigner function of NGBS. Specifically, Wigner function of NGBS are computed for
various choices of the state parameters. The obtained Wigner functions for
the NGBS are shown in Figs. \ref{fig:Wigner-GBS}(a)-(d). Clearly,
there exist negative regions of Wigner function in all the figures
and that establishes that the NGBS studied
here are nonclassical. However, neither negativity of the Wigner function nor the nonclassicality witnesses studied till now  provide any quantitative measure of the nonclassicality. The same
may be obtained using nonclassical volume which is described 
in the following subsection.

\subsection{Nonclassical volume}

From the negativity of the Wigner function, we have obtained the signature
of nonclassicality in NGBS. However, the amount of nonclassicality
is not yet investigated. There exist several quantitative measures
of nonclassicality. Here we may use the most convenient quantitative
measure of nonclassicality which is known as nonclassical volume.
The nonclassical volume as a measure of quantumness was first introduced
by Kenfack and Zyczkowski \cite{kenfack2004negativity} in 2004. In
this particular measure, the volume of the negative part of the Wigner
function is considered as the measure of nonclassicality. To be precise,
the negative volume associated with a quantum state $|\psi\rangle$
is

\[
\delta(\psi)=\int\int\left|W_{\psi}\left(p',q\right)\right|dqdp'-1,
\]
where $W_{\psi}\left(p',q\right)$ is the Wigner function of a quantum
state $|\psi\rangle.$ A non-zero values of $\delta(\psi)$ indicates
nonclassical state. As we have a compact expression for the Wigner function,
we can use the same to obtain $\delta(\psi)$ for various choices
of parameters and investigate how nonclassical volume (or the amount
of quantumness) varies with the change of a particular parameter.
For example, in Table \ref{tab:Nonclassical-volume}, we have
shown the variation of $\delta(\psi)$ with probability $p$ for the
fixed values of $M$ in NGBS and it is found that the amount of nonclassicality
increases with $p$, which was also indicated by the nonclassicality witnesses, studied in this paper. 

\begin{table}
\begin{centering}
\begin{tabular}{ccc}
\hline 
Serial Number & Probability $(p)$ & Nonclassical volume $(\delta(\psi))$\tabularnewline
\hline 
\hline 
1. & 0.2 & 0.166724\tabularnewline
2. & 0.4 & 0.244092\tabularnewline
3. & 0.6 & 0.324178\tabularnewline
4. & 0.8 & 0.416412\tabularnewline
\hline 
\end{tabular}
\par\end{centering}
\caption{\label{tab:Nonclassical-volume}Nonclassical volume for NGBS with
$q=0.5$ and $M=25$.}
\end{table}

\subsection{Optical tomogram \label{subsec:Optical-tomogram}}

\begin{figure}
\centering{}

\subfigure[]{\includegraphics[scale=0.6]{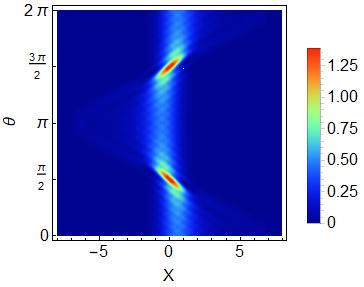}}\quad \subfigure[]{\includegraphics[scale=0.6]{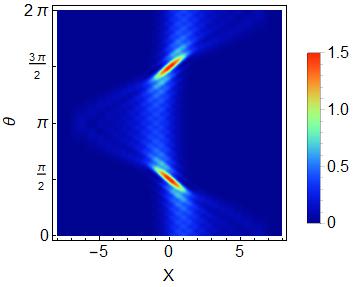}}\\
\subfigure[]{\includegraphics[scale=0.6]{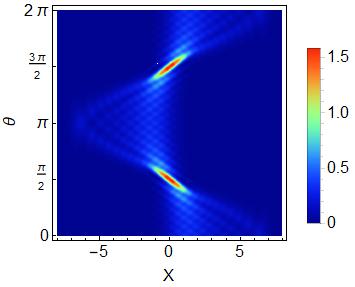}}\quad \subfigure[]{\includegraphics[scale=0.6]{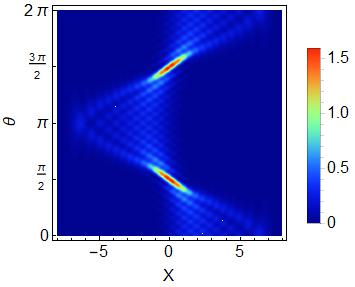}}
\caption{\label{fig:Tomograms}(Color online) Tomograms for the NGBS for (a) $p=0.2$, (b)
$p=0.4$, (c) $p=0.6$ and (d) $p=0.8$, respectively. In all figures
$M=25$ and $q=0.5$.}
\end{figure}

There exist some proposals for the direct measurement of Wigner function
\cite{banaszek1999direct,bertet2002direct,shalibo2013direct}. However,
in general, due to its probabilistic nature, the direct measurement of Wigner function is not possible, so we may experimentally measure only by processing of data. Interestingly, tomogram gives the probabilistic description of the quantum state which is accessible for  direct measurement. Therefore, Wigner function can be measured by using optical homodyne tomomgraphy  \cite{smithey1993measurement}. For any quantum state
$|\psi\rangle$, the optical tomogram $w_{|\psi\rangle}(X,\theta)$
is the marginal distribution of the field quadrature component $X$
with a rotation by an angle $\theta$ in the quadrature phase space
\cite{filippov2011optical}. Mathematically, optical tomogram can be obtained from the Radon transform of the Wigner function. 
Of late, Filippov
and Man'ko \cite{filippov2011optical,pathak2014wigner} have reported
the following closed form analytic expression for the optical tomogram of finite dimensional FSS (\ref{eq:quantum state1}):

\begin{equation}
\begin{array}{lcl}
w_{|\psi\rangle}(X,\theta) & = & \frac{e^{-X^{2}}}{\sqrt{\pi}}\left[\sum_{n=0}^{N}\frac{|c_{n}|^{2}}{2^{n}n!}H_{n}^{2}(X)+\sum_{n<k}\frac{|c_{n}||c_{k}|\cos\left(\left(n-k\right)\theta-\left(\phi_{n}-\phi_{k}\right)\right)}{\sqrt{2^{n+k-2}n!k!}}H_{n}(X)H_{k}(X)\right]\end{array}\label{eq:optical tomogram}
\end{equation}
where $c_{j}=|c_{j}|e^{i\phi_{j}}$ and $H_{j}$ is the Hermite polynomial
of degree $j$. The optical tomogram of NGBS has been calculated by using
Eqs. (\ref{eq:NGBS-coefficient}) and (\ref{eq:optical tomogram}).
The results are depicted in Figs. \ref{fig:Tomograms}(a)-(d). The
Wigner function for the NGBS can be experimentally verified by optical
homodyne tomography technique where the expected tomogram can be obtained
theoretically from Eq. (\ref{eq:optical tomogram}).

\section{Conclusion\label{sec:Conclusion} }

Finite dimensional quantum states or qudits have drawn  attention of the scientific community as these states have been found to be useful in performing various tasks related to quantum information processing. Construction of such states have become feasible with the  rapid development of  quantum state
engineering. One such interesting finite dimensional quantum state is NGBS which reduces to various states of interest in different limits. Nonclassical properties of this state are studied here with a focus on higher-order
nonclassical properties. The analytically obtained expressions for various witnesses of nonclassicality are plotted to establish that NGBS is highly nonclassical and thus can be used for various tasks where nonclassical states are essential.  Specifically, the presence of higher-order nonclassical
properties (HOA, HOS and HOSPS) in NGBS are shown here for the first time. 

Physically, among various intermediate states and the states from the family of binomial and generalized binomial states,
NGBS is of particular interest because it is useful when the photon
emitting probabilities of a system are not equal hence it is applicable in a more generalized procedure. Physical objective of this generalization procedure is to model a complex physical system such as a molecular system which have different photon emitting probabilities  from different energy levels of an excited molecule during any nonlinear procedure. 
Therefore, this scenario appears to be obvious during
any nonlinear process and its experimental realization \cite{fan1999new}. Thus, the focus of this paper (investigation on NGBS) seems justified from this practical scenario. To investigate
nonclassical witnesses and measures in this generalized engineered
state, we have used a set of moment-based criteria and also Wigner quasi-probability distribution.  Specifically, the
moment-based criteria used here to analyze nonclassical features include the criteria for 
lower- and higher-order antibunching, HOSPS, Hong Mandel HOS, Hillery
HOS,  Agarwal-Tara criterion and
Vogel criterion. The moment-based criteria and quasi-probability distribution,
investigated here, have revealed a variety of nonclassical properties
of NGBS. In most of the cases, depth of nonclassicality witness is found to increase
with dimension but it decreases with  the parameter $q$ which is introduced for different
photon emitting probabilities in NGBS. However, in a few occasions, namely in the context of  HOS Hillery type
and Vogel criterion,  the results are not always consistent with other criteria. In the case of Vogel's criterion, it is observed that the depth of the nonclassical
witness increases with dimension $M$ only for the higher values of probability and this result is not consistent for the lower value of probability (cf. Figs. \ref{fig.vogel}(b) and (c)). On the other hand, in the case of HOS the depth of the inequality increases in the negative region with increase of the parameter $q$ for  higher values of probability but it decreases with $q$  for lower values of probability.  Hence for lower values of probability $p$,  the result  is consistent with other criteria (cf. Fig.\ref{fig.HOS}(d)). The Wigner quasi-probability distribution
of the NGBS are reported and  the amount of nonclassicality is computed by computing nonclassical volume, which is nothing but the volume of the negative part of the Wigner function.
The nonclassical volume (thus the amount of nonclassicality) of NGBS  is found to increase with probability
$p$. In short, it is established that for higher value of the
probability, the amount of nonclassicality in NGBS is high. The nature
of optical tomograms for NGBS are explored and reported for different probability. Tomograms can be obtained experimentally and the same may be used to obtain Wigner function with the help of Radon transform. Keeping all these facts in mind, we conclude the paper with the hope that the present observation will be useful
in quantum optics and quantum information processing, specifically, in the experimental scenario
when the photon emitting probability is not equal. 

\subsection*{Acknowledgment: }

A.P. and N.A. thank the Department of Science and Technology (DST),
India, for support provided through the DST project No. EMR/2015/000393.

\bibliographystyle{unsrt}
\bibliography{article}

\begin{thebibliography}{10}

\bibitem{marchiolli2004engineering}
Marcelo~A Marchiolli and Wagner~Duarte Jos{\'e}.
\newblock Engineering superpositions of displaced number states of a trapped
  ion.
\newblock {\em Physica A: Statistical Mechanics and its Applications},
  337(1-2):89--108, 2004.

\bibitem{miranowicz2004dissipation}
Adam Miranowicz and Wies{\l}aw Leo{\'n}ski.
\newblock Dissipation in systems of linear and nonlinear quantum scissors.
\newblock {\em Journal of Optics B: Quantum and Semiclassical Optics},
  6(3):S43, 2004.

\bibitem{vogel1993quantum}
K~Vogel, VM~Akulin, and WP~Schleich.
\newblock Quantum state engineering of the radiation field.
\newblock {\em Physical Review Letters}, 71(12):1816, 1993.

\bibitem{sperling2014quantum}
J~Sperling, W~Vogel, and GS~Agarwal.
\newblock Quantum state engineering by click counting.
\newblock {\em Physical Review A}, 89(4):043829, 2014.

\bibitem{nielsen2002quantum}
Michael~A Nielsen and Isaac Chuang.
\newblock Quantum computation and quantum information, 2002.

\bibitem{pathak2013elements}
Anirban Pathak.
\newblock {\em Elements of quantum computation and quantum communication}.
\newblock CRC Press, 2013.

\bibitem{barnett2018statistics}
Stephen~M Barnett, Gergely Ferenczi, Claire~R Gilson, and Fiona~C Speirits.
\newblock Statistics of photon-subtracted and photon-added states.
\newblock {\em Physical Review A}, 98(1):013809, 2018.

\bibitem{verma2010generalized}
Amit Verma and Anirban Pathak.
\newblock Generalized structure of higher order nonclassicality.
\newblock {\em Physics Letters A}, 374(8):1009--1020, 2010.

\bibitem{pathak2014wigner}
Anirban Pathak and J~Banerji.
\newblock Wigner distribution, nonclassicality and decoherence of generalized
  and reciprocal binomial states.
\newblock {\em Physics Letters A}, 378(3):117--123, 2014.

\bibitem{verma2008higher}
Amit Verma, Navneet~K Sharma, and Anirban Pathak.
\newblock Higher order antibunching in intermediate states.
\newblock {\em Physics Letters A}, 372(34):5542--5551, 2008.

\bibitem{fu1997hypergeometric}
Hong-Chen Fu and Ryu Sasaki.
\newblock Hypergeometric states and their nonclassical properties.
\newblock {\em Journal of Mathematical Physics}, 38(5):2154--2166, 1997.

\bibitem{moussa1998generation}
MHY Moussa and B~Baseia.
\newblock Generation of the reciprocal-binomial state.
\newblock {\em Physics Letters A}, 238(4-5):223--226, 1998.

\bibitem{harrow2017quantum}
Aram~W Harrow and Ashley Montanaro.
\newblock Quantum computational supremacy.
\newblock {\em Nature}, 549(7671):203, 2017.

\bibitem{neill2018blueprint}
C~Neill, P~Roushan, K~Kechedzhi, S~Boixo, SV~Isakov, V~Smelyanskiy, A~Megrant,
  B~Chiaro, A~Dunsworth, K~Arya, et~al.
\newblock A blueprint for demonstrating quantum supremacy with superconducting
  qubits.
\newblock {\em Science}, 360(6385):195--199, 2018.

\bibitem{thapliyal2014higher}
Kishore Thapliyal, Anirban Pathak, Biswajit Sen, and Jan Pe{\v{r}}ina.
\newblock Higher-order nonclassicalities in a codirectional nonlinear optical
  coupler: Quantum entanglement, squeezing, and antibunching.
\newblock {\em Physical Review A}, 90(1):013808, 2014.

\bibitem{thapliyal2014nonclassical}
Kishore Thapliyal, Anirban Pathak, Biswajit Sen, and Jan Perina.
\newblock Nonclassical properties of a contradirectional nonlinear optical
  coupler.
\newblock {\em Physics Letters A}, 378(46):3431--3440, 2014.

\bibitem{giri2014single}
Sandip~Kumar Giri, Biswajit Sen, CH~Raymond Ooi, and Anirban Pathak.
\newblock Single-mode and intermodal higher-order nonclassicalities in two-mode
  bose-einstein condensates.
\newblock {\em Physical Review A}, 89(3):033628, 2014.

\bibitem{alam2017lower}
Nasir Alam, Kishore Thapliyal, Anirban Pathak, Biswajit Sen, Amit Verma, and
  Swapan Mandal.
\newblock Lower-and higher-order nonclassicality in a bose-condensed
  optomechanical-like system and a fabry-perot cavity with one movable mirror:
  squeezing, antibunching and entanglement.
\newblock {\em arXiv preprint arXiv:1708.03967}, 2017.

\bibitem{alam2015approximate}
Nasir Alam, Swapan Mandal, and Patrik {\"O}hberg.
\newblock Approximate analytical solutions of a pair of coupled anharmonic
  oscillators.
\newblock {\em Journal of Physics B: Atomic, Molecular and Optical Physics},
  48(4):045503, 2015.

\bibitem{alam2016nonclassical}
Nasir Alam and Swapan Mandal.
\newblock Nonclassical properties of coherent light in a pair of coupled
  anharmonic oscillators.
\newblock {\em Optics Communications}, 359:221--233, 2016.

\bibitem{alam2018higher}
Nasir Alam, Amit Verma, and Anirban Pathak.
\newblock Higher-order nonclassicalities of finite dimensional coherent states:
  A comparative study.
\newblock {\em Physics Letters A}, 382(28):1842--1851, 2018.

\bibitem{Meher2018}
Nilakantha Meher and S.~Sivakumar.
\newblock Number state filtered coherent states.
\newblock {\em Quantum Information Processing}, 17(9):233, Jul 2018.

\bibitem{malpani2018lower}
Priya Malpani, Nasir Alam, Kishore Thapliyal, Anirban Pathak, V~Narayanan, and
  Subhashish Banerjee.
\newblock Lower-and higher-order nonclassical properties of photon added and
  subtracted displaced fock states.
\newblock {\em arXiv preprint arXiv:1808.01458}, 2018.

\bibitem{Alam2018}
Nasir Alam, Kathakali Mandal, and Anirban Pathak.
\newblock Higher-order nonclassical properties of a shifted symmetric cat state
  and a one-dimensional continuous superposition of coherent states.
\newblock {\em International Journal of Theoretical Physics},
  57(11):3443--3456, Nov 2018.

\bibitem{abbott2016bp}
BP~Abbott.
\newblock Bp abbott, r. abbott, td abbott, mr abernathy, f. acernese, k.
  ackley, c. adams, t. adams, p. addesso, rx adhikari et al., phys. rev. lett.
  116, 061102 (2016).
\newblock {\em Phys. Rev. Lett.}, 116:061102, 2016.

\bibitem{abbott2016gw151226}
BP~Abbott, R~Abbott, TD~Abbott, MR~Abernathy, F~Acernese, K~Ackley, C~Adams,
  T~Adams, P~Addesso, RX~Adhikari, et~al.
\newblock Gw151226: Observation of gravitational waves from a 22-solar-mass
  binary black hole coalescence.
\newblock {\em Physical Review Letters}, 116(24):241103, 2016.

\bibitem{gottesman2003secure}
Daniel Gottesman and John Preskill.
\newblock Secure quantum key distribution using squeezed states.
\newblock In {\em Quantum Information with Continuous Variables}, pages
  317--356. Springer, 2003.

\bibitem{cerf2001quantum}
Nicolas~J Cerf, Marc Levy, and Gilles Van~Assche.
\newblock Quantum distribution of gaussian keys using squeezed states.
\newblock {\em Physical Review A}, 63(5):052311, 2001.

\bibitem{madsen2012continuous}
Lars~S Madsen, Vladyslav~C Usenko, Mikael Lassen, Radim Filip, and Ulrik~L
  Andersen.
\newblock Continuous variable quantum key distribution with modulated entangled
  states.
\newblock {\em Nature communications}, 3:1083, 2012.

\bibitem{weedbrook2012gaussian}
Christian Weedbrook, Stefano Pirandola, Ra{\'u}l Garc{\'\i}a-Patr{\'o}n,
  Nicolas~J Cerf, Timothy~C Ralph, Jeffrey~H Shapiro, and Seth Lloyd.
\newblock Gaussian quantum information.
\newblock {\em Reviews of Modern Physics}, 84(2):621, 2012.

\bibitem{bennett1993teleporting}
Charles~H Bennett, Gilles Brassard, Claude Cr{\'e}peau, Richard Jozsa, Asher
  Peres, and William~K Wootters.
\newblock Teleporting an unknown quantum state via dual classical and
  einstein-podolsky-rosen channels.
\newblock {\em Physical Review Letters}, 70(13):1895, 1993.

\bibitem{ekert1991quantum}
Artur~K Ekert.
\newblock Quantum cryptography based on bell's theorem.
\newblock {\em Physical Review Letters}, 67(6):661, 1991.

\bibitem{bennett1992quantum}
Charles~H Bennett, Gilles Brassard, and Artur~K Ekert.
\newblock Quantum cryptography.
\newblock {\em Scientific American}, 267(4):50--57, 1992.

\bibitem{pathak2010recent}
Anirban Pathak and Amit Verma.
\newblock Recent developments in the study of higher order nonclassical states.
\newblock {\em Indian Journal of Physics}, 84(8):1005--1019, 2010.

\bibitem{shukla2014protocols}
Chitra Shukla, Nasir Alam, and Anirban Pathak.
\newblock Protocols of quantum key agreement solely using bell states and bell
  measurement.
\newblock {\em Quantum Information Processing}, 13(11):2391--2405, 2014.

\bibitem{stoler1985binomial}
D~Stoler, BEA Saleh, and MC~Teich.
\newblock Binomial states of the quantized radiation field.
\newblock {\em Optica Acta: International Journal of Optics}, 32(3):345--355,
  1985.

\bibitem{fan1999new}
Hong-Yi Fan et~al.
\newblock New generalized binomial states of the quantized radiation field.
\newblock {\em Physics Letters A}, 264(2-3):154--161, 1999.

\bibitem{agarwal_2012}
Girish~S. Agarwal.
\newblock {\em Quantum Optics}.
\newblock Cambridge University Press, 2012.

\bibitem{agarwal1992negative}
GS~Agarwal.
\newblock Negative binomial states of the field-operator representation and
  production by state reduction in optical processes.
\newblock {\em Physical Review A}, 45(3):1787, 1992.

\bibitem{barnett1998negative}
Stephen~M Barnett.
\newblock Negative binomial states of the quantized radiation field.
\newblock {\em Journal of Modern Optics}, 45(10):2201--2205, 1998.

\bibitem{franco2009quantum}
Rosario~Lo Franco, Giuseppe Compagno, Antonino Messina, and Anna Napoli.
\newblock Quantum computation with generalized binomial states in cavity
  quantum electrodynamics.
\newblock {\em International Journal of Quantum Information},
  7(supp01):155--162, 2009.

\bibitem{lee1990higher}
Ching~Tsung Lee.
\newblock Higher-order criteria for nonclassical effects in photon statistics.
\newblock {\em Physical Review A}, 41(3):1721, 1990.

\bibitem{hong1985generation}
CK~Hong and Li~Mandel.
\newblock Generation of higher-order squeezing of quantum electromagnetic
  fields.
\newblock {\em Physical Review A}, 32(2):974, 1985.

\bibitem{giri2017nonclassicality}
Sandip~Kumar Giri, Kishore Thapliyal, Biswajit Sen, and Anirban Pathak.
\newblock Nonclassicality in an atom--molecule bose--einstein condensate:
  Higher-order squeezing, antibunching and entanglement.
\newblock {\em Physica A: Statistical Mechanics and its Applications},
  466:140--152, 2017.

\bibitem{allevi2012high}
Alessia Allevi, Stefano Olivares, and Maria Bondani.
\newblock High-order photon-number correlations: a resource for
  characterization and applications of quantum states.
\newblock {\em International Journal of Quantum Information}, 10(08):1241003,
  2012.

\bibitem{allevi2012measuring}
Alessia Allevi, Stefano Olivares, and Maria Bondani.
\newblock Measuring high-order photon-number correlations in experiments with
  multimode pulsed quantum states.
\newblock {\em Physical Review A}, 85(6):063835, 2012.

\bibitem{arrazola2017quantum}
Juan~Miguel Arrazola, Patrick Rebentrost, and Christian Weedbrook.
\newblock Quantum supremacy and high-dimensional integration.
\newblock {\em arXiv preprint arXiv:1712.07288}, 2017.

\bibitem{valverde2003generation}
C~Valverde, AT~Avelar, B~Baseia, and JMC Malbouisson.
\newblock Generation of the reciprocal-binomial state for optical fields.
\newblock {\em Physics Letters A}, 315(3-4):213--218, 2003.

\bibitem{franco2006single}
R~Lo Franco, Giuseppe Compagno, A~Messina, and A~Napoli.
\newblock Single-shot generation and detection of a two-photon generalized
  binomial state in a cavity.
\newblock {\em Physical Review A}, 74(4):045803, 2006.

\bibitem{franco2010efficient}
Rosario~Lo Franco, Giuseppe Compagno, Antonino Messina, and Anna Napoli.
\newblock Efficient generation of n-photon binomial states and their use in
  quantum gates in cavity qed.
\newblock {\em Physics Letters A}, 374(22):2235--2242, 2010.

\bibitem{zavatta2004quantum}
Alessandro Zavatta, Silvia Viciani, and Marco Bellini.
\newblock Quantum-to-classical transition with single-photon-added coherent
  states of light.
\newblock {\em science}, 306(5696):660--662, 2004.

\bibitem{an2002multimode}
Nguyen~Ba An.
\newblock Multimode higher-order antibunching and squeezing in trio coherent
  states.
\newblock {\em Journal of Optics B: Quantum and Semiclassical Optics},
  4(3):222, 2002.

\bibitem{pathak2006control}
A~Pathak and ME~Garcia.
\newblock Control of higher order antibunching.
\newblock {\em Applied Physics B}, 84(3):479--484, 2006.

\bibitem{hillery1987amplitude}
Mark Hillery.
\newblock Amplitude-squared squeezing of the electromagnetic field.
\newblock {\em Physical Review A}, 36(8):3796, 1987.

\bibitem{agarwal1992nonclassical}
GS~Agarwal and K~Tara.
\newblock Nonclassical character of states exhibiting no squeezing or
  sub-poissonian statistics.
\newblock {\em Physical Review A}, 46(1):485, 1992.

\bibitem{shchukin2005nonclassical}
Evgeny~V Shchukin and Werner Vogel.
\newblock Nonclassical moments and their measurement.
\newblock {\em Physical Review A}, 72(4):043808, 2005.

\bibitem{moya1993series}
H{\'e}ctor Moya-Cessa and Peter~L Knight.
\newblock Series representation of quantum-field quasiprobabilities.
\newblock {\em Physical Review A}, 48(3):2479, 1993.

\bibitem{kenfack2004negativity}
Anatole Kenfack and Karol {\.Z}yczkowski.
\newblock Negativity of the wigner function as an indicator of
  non-classicality.
\newblock {\em Journal of Optics B: Quantum and Semiclassical Optics},
  6(10):396, 2004.

\bibitem{banaszek1999direct}
K~Banaszek, C~Radzewicz, K~W{\'o}dkiewicz, and JS~Krasi{\'n}ski.
\newblock Direct measurement of the wigner function by photon counting.
\newblock {\em Physical Review A}, 60(1):674, 1999.

\bibitem{bertet2002direct}
Patrice Bertet, Alexia Auffeves, Paolo Maioli, Stefano Osnaghi, Tristan
  Meunier, Michel Brune, Jean-Michel Raimond, and Serge Haroche.
\newblock Direct measurement of the wigner function of a one-photon fock state
  in a cavity.
\newblock {\em Physical Review Letters}, 89(20):200402, 2002.

\bibitem{shalibo2013direct}
Yoni Shalibo, Roy Resh, Ofer Fogel, David Shwa, Radoslaw Bialczak, John~M
  Martinis, and Nadav Katz.
\newblock Direct wigner tomography of a superconducting anharmonic oscillator.
\newblock {\em Physical Review Letters}, 110(10):100404, 2013.

\bibitem{smithey1993measurement}
DT~Smithey, M~Beck, Michael~G Raymer, and A~Faridani.
\newblock Measurement of the wigner distribution and the density matrix of a
  light mode using optical homodyne tomography: Application to squeezed states
  and the vacuum.
\newblock {\em Physical Review Letters}, 70(9):1244, 1993.

\bibitem{filippov2011optical}
Sergey~N Filippov and Vladimir~I Man'ko.
\newblock Optical tomography of fock state superpositions.
\newblock {\em Physica Scripta}, 83(5):058101, 2011.

\bibitem{gradshteyn2001table}
I.S. Gradshteyn and I.M. Ryzhik.
\newblock {\em Table of Integrals, Series, and Products}.
\newblock Academic Press, 2001.

\end{thebibliography}

\newpage
\appendix
\section*{APPENDIX: A}  

\renewcommand{\theequation}{A-\arabic{equation}}    
  \setcounter{equation}{0}  
  
\subsection*{Detailed derivation of Wigner function}

Fock state $|n\rangle$ can be written in position space as 
\begin{equation}
|n\rangle=\phi_{n}(x)=b_{n}{\rm e}^{-\frac{x^{2}}{2}}H_{n}(x),\label{eq:fock-state in corrdinate space}
\end{equation}
where $b_{n}=\frac{1}{\pi^{\frac{1}{4}}[2^{n}n!]^{\frac{1}{2}}}$,
$\hbar=1$ and $H_{n}(x)$ is the Hermite polynomial. Consequently,
we can express the FSS (\ref{eq:quantum state1}) as 
\[
\psi(x)=\sum_{n=0}^{N}c_{n}\phi_{n}(x).
\]
 Therefore, the Wigner function of an arbitrary FSS in the integral form
is 
\begin{equation}
\begin{array}{lcl}
W(x,p') & = & \frac{1}{\pi}\intop_{-\infty}^{\infty}\psi^{*}(x+y)\psi(x-y)\exp\left(2ip'y\right)dy\\
 & = & \frac{1}{\pi}\sum_{n,n^{\prime}=0}^{N}c_{n}^{*}c_{n^{\prime}}\intop_{-\infty}^{\infty}\phi_{n}^{*}(x+y)\phi_{n^{\prime}}(x-y)\exp(2ip'y)dy\\
 & = & \frac{1}{\pi}\sum_{n,n^{\prime}=0}^{N}c_{n}^{*}c_{n^{\prime}}W_{nn^{\prime}}(x,p'),
\end{array}\label{eq:wigner-our}
\end{equation}
 where 
\begin{equation}
W_{nn^{\prime}}(x,p')=\intop_{-\infty}^{\infty}\phi_{n}^{*}(x+y)\phi_{n^{\prime}}(x-y)\exp(2ip'y)dy.\label{eq:wnn1}
\end{equation}
 Using (\ref{eq:fock-state in corrdinate space}), we can write
\begin{equation}
\begin{array}{lcl}
W_{nn^{\prime}}(x,p') & = & b_{n}^{*}b_{n^{\prime}}\intop_{-\infty}^{\infty}{\rm e}^{-\frac{(x+y)^{2}}{2}}H_{n}(x+y){\rm e}^{-\frac{(x-y)^{2}}{2}}H_{n^{\prime}}(x-y){\rm e}^{2ip'y}dy\\
 & = & b_{n}^{*}b_{n^{\prime}}\intop_{-\infty}^{\infty}H_{n}(x+y)H_{n^{\prime}}(x-y){\rm e}^{-x^{2}-y^{2}+2ip'y}dy\\
 & = & b_{n}^{*}b_{n^{\prime}}{\rm e}^{-x^{2}}\intop_{-\infty}^{\infty}H_{n}(x+y)H_{n^{\prime}}(x-y){\rm e}^{-(y-ip')^{2}-p'^{2}}dy\\
 & = & b_{n}^{*}b_{n^{\prime}}{\rm e}^{-\left(x^{2}+p'^{2}\right)}I_{nn^{\prime}},
\end{array}\label{eq:wnn2}
\end{equation}
where 
\begin{equation}
I_{nn^{\prime}}=\intop_{-\infty}^{\infty}H_{n}(x+y)H_{n^{\prime}}(x-y){\rm e}^{-(y-ip')^{2}}dy.\label{eq:inn}
\end{equation}
 Writing $y-ip'=y^{\prime}$ in (\ref{eq:inn}) we obtain
\begin{equation}
\begin{array}{lcl}
I_{nn^{\prime}} & = & \intop_{-\infty}^{\infty}H_{n}(x+y^{\prime}+ip')H_{n^{\prime}}(x-y^{\prime}-ip'){\rm e}^{-(y^{\prime})^{2}}dy^{\prime}\\
 & = & (-1)^{n^{\prime}}\intop_{-\infty}^{\infty}H_{n}(y^{\prime}+(x+ip'))H_{n^{\prime}}(y^{\prime}+(ip'-x)){\rm e}^{-(y^{\prime})^{2}}dy^{\prime}\\
 & = & (-1)^{n^{\prime}}2^{n^{\prime}}\pi^{\frac{1}{2}}n!(ip'-x)^{n^{\prime}-n}L_{n}^{n^{\prime}-n}\left(-2y(ip'-x)\right),\,\,\,\,\,n\leq n^{\prime}.
\end{array}\label{eq:Inn2}
\end{equation}
In the last step, we have used the identity 7.377 of \cite{gradshteyn2001table}.
Now substituting (\ref{eq:wnn2}) and (\ref{eq:Inn2}) in (\ref{eq:wigner-our})
we obtain 
\begin{equation}
W(x,p')=\frac{1}{\pi^{\frac{1}{2}}}\sum_{n,n^{\prime}=0}^{N}c_{n}^{*}c_{n^{\prime}}b_{n}^{*}b_{n^{\prime}}{\rm e}^{-\left(x^{2}+p'^{2}\right)}(-1)^{n^{\prime}}2^{n^{\prime}}n!(ip'-x)^{n^{\prime}-n}L_{n}^{n^{\prime}-n}\left(-2y(ip'-x)\right),\,\,\,\,\,n\leq n^{\prime}.\label{Apprndix-wigner}
\end{equation}

\end{document}